\documentclass[aps,prd,onecolumn,groupedaddress,showpacs,nofootinbib,amssymb]{revtex4}
\usepackage[T1]{fontenc}
\usepackage[latin1]{inputenc}
\usepackage{graphicx}
\usepackage[english]{babel}
\usepackage{amsmath}
\usepackage{amssymb}
\usepackage{amsfonts}

\begin{document}

\def\pp{{\, \mid \hskip -1.5mm =}}
\def\cL{{\cal L}}
\def\be{\begin{equation}}
\def\ee{\end{equation}}
\def\bea{\begin{eqnarray}}
\def\eea{\end{eqnarray}}
\def\beq{\begin{eqnarray}}
\def\eeq{\end{eqnarray}}
\def\tr{{\rm tr}\, }
\def\nn{\nonumber \\}
\def\e{{\rm e}}

\title{Cosmological reconstruction of realistic modified $F(R)$ gravities}

\author{Shin'ichi Nojiri$^1$, Sergei D. Odintsov$^{2,3}$\footnote{
Also at Center of Theor. Physics, TSPU, Tomsk}, and Diego S\'{a}ez-G\'{o}mez$^{3}$}
\affiliation{
$^1$Department of Physics, Nagoya University, Nagoya 464-8602, Japan \\
$^2$Instituci\`{o} Catalana de Recerca i Estudis Avan\c{c}ats (ICREA),
Barcelona \\
$^3$ Institut de Ciencies de l'Espai (IEEC-CSIC),
Campus UAB, Facultat de Ciencies, Torre C5-Par-2a pl, E-08193 Bellaterra (Barcelona), Spain}

\begin{abstract}

The cosmological reconstruction scheme for modified $F(R)$ gravity is
developed in terms of e-folding (or, redshift). It is demonstrated how any
FRW cosmology may emerge from specific $F(R)$ theory. The specific
examples of well-known cosmological evolution are reconstructed, including
$\Lambda$CDM cosmology, deceleration with transition to phantom
superacceleration era which may develop singularity or be transient.
The application of this scheme to viable $F(R)$ gravities unifying
inflation with dark energy era is proposed. The additional reconstruction
of such models leads to non-leading gravitational correction mainly
relevant at the early/late universe and helping to pass the cosmological
bounds (if necessary). It is also shown how cosmological reconstruction
scheme may be generalized in the presence of scalar field.

\end{abstract}


\maketitle

\section{Introduction}

Modified gravity approach suggests the gravitational alternative for
unified description of inflation, dark energy and dark matter without the
need to introduce by hands the inflaton and extra dark components.
Moreover, the
easy explanation of inflationary or dark energy phase in such scenario
follows: the corresponding era is emerging due to dominance of the
specific gravitational sector in the course of the universe expansion.
In other words, the early-time and late-time acceleration is governed by
the universe expansion within the specific modified gravity theory.
Special interest in this gravitational paradigm for the description of the
universe evolution is related with $F(R)$ gravity (for a general review,
see \cite{review}) due to its quite simple structure if compare with more
general modified gravity which includes all curvature invariants as well
as non-local terms. Nevertheless, even in frames of $F(R)$ gravity the
background evolution (due to high non-linearity of the problem) is
often non-explicit and/or
non-analytic process. From another side, any realistic modified gravity
should pass not only the local tests but also the
observational cosmological bounds. To comply with cosmological bounds, the
reconstruction program in any modified gravity has been 
developed \cite{reconstruction1}.

The cosmological reconstruction of $F(R)$ gravity has been considered in
refs.\cite{reconstruction1,reconstruction2,reconstruction3,bamba}. It
turns out that in most cases this reconstruction is done in the
presence of the auxiliary scalar which may be excluded at the final step so
that any FRW cosmology may be realized within specific reconstructed $F(R)$
gravity. However, the weak point of so developed reconstruction scheme is
that the final function $F(R)$ represents usually some polynomial in the
positive/negative powers of scalar curvature. On the same time, the viable
models have strongly non-linear structure.

In the present paper we develop the new scheme for cosmological
reconstruction of $F(R)$ gravity in terms of e-folding (or, redshift $z$) so
that there is no
need to use more complicated formulation with auxiliary
scalar \cite{reconstruction1,reconstruction2,bamba}. Using such technique
the number of examples are presented where $F(R)$ gravity is reconstructed
so that it gives the well-known cosmological evolution: $\Lambda$CDM
epoch, deceleration/acceleration epoch which is equivalent to presence of
 phantom and non-phantom matter, late-time acceleration with the
crossing of phantom-divide line, transient phantom epoch and oscillating
universe. It is shown that some generalization of such technique for
viable $F(R)$ gravity is possible, so that local tests are usually
satisfied. In this way, modified gravity unifying inflation,
radiation/matter dominance and dark energy eras may be further
reconstructed in the early or in the late universe so that the future
evolution may be different.
This opens the way to non-linear reconstruction of realistic $F(R)$ gravity.
Moreover, it is demonstrated that cosmological reconstruction
of viable modified gravity may help in the formulation of non-singular
models in finite-time future. The reconstruction suggests the way to change
some cosmological predictions of the theory in the past or in the future
so that it becomes easier to pass the available observational data.
Finally, we show that our method works also for modified gravity with
scalar theory and any requested cosmology may be realized within such
theory too.

\section{Cosmological reconstruction of modified $F(R)$ gravity}

Let us demonstrate that any FRW cosmology may be realized in specific
$F(R)$ gravity.
The starting action of the $F(R)$ gravity (for general review, 
see \cite{review}) is given by
\be
\label{Hm0}
S= \int d^4x \sqrt{-g} \left(\frac{F(R)}{2\kappa^2} + \mathcal{L}_\mathrm{matter} \right)\ .
\ee
The field equation corresponding to the first FRW equation is:
\be
\label{Hm1}
0 = - \frac{F(R)}{2} + 3 \left( H^2 + \dot H\right) F'(R)
 - 18 (\left(4 H^2 \dot H + H\ddot H\right) F''(R) + \kappa^2 \rho\ .
\ee
with $R=6\dot H + 12 H^2$. We now rewrite Eq.(\ref{Hm1}) by using a new variable
(which is often called e-folding) instead of the cosmological time $t$,
$N=\ln \frac{a}{a_0}$.
The variable $N$ is related with the redshift $z$ by
$\e^{-N}=\frac{a_0}{a} = 1 + z$.
Since $\frac{d}{dt} = H \frac{d}{dN}$ and therefore
$\frac{d^2}{dt^2} = H^2 \frac{d^2}{dN^2} + H \frac{dH}{dN} \frac{d}{dN}$,
one can rewrite (\ref{Hm1}) by
\be
\label{RZ4}
0 = - \frac{F(R)}{2} + 3 \left( H^2 + H H'\right) F'(R)
 - 18 (\left(4 H^3 H' + H^2 \left(H'\right)^2 + H^3 H''\right) F''(R) + \kappa^2 \rho\ .
\ee
Here $H'\equiv dH/dN$ and $H''\equiv d^2 H/dN^2$.
If the matter energy density $\rho$ is given by a sum of the fluid
densities with constant EoS
parameter $w_i$, we find
\be
\label{RZ6}
\rho=\sum_i \rho_{i0} a^{-3(1+w_i)} = \sum_i \rho_{i0} a_0^{-3(1+w_i)} \e^{-3(1+w_i)N}\ .
\ee
Let the Hubble rate is given in terms of $N$ via the function $g(N)$ as
\be
\label{RZ7}
H=g(N) = g \left(- \ln\left(1+z\right)\right)\ .
\ee
Then scalar curvature takes the form: $R = 6 g'(N) g(N) + 12 g(N)^2$,
which could be solved with respect to $N$ as
$N=N(R)$. Then by using (\ref{RZ6}) and (\ref{RZ7}), one can rewrite
(\ref{RZ4}) as
\bea
\label{RZ9}
0 &=& -18 \left(4g\left(N\left(R\right)\right)^3 g'\left(N\left(R\right)\right)
+ g\left(N\left(R\right)\right)^2 g'\left(N\left(R\right)\right)^2
+ g\left(N\left(R\right)\right)^3g''\left(N\left(R\right)\right)\right) \frac{d^2 F(R)}{dR^2} \nn
&& + 3 \left( g\left(N\left(R\right)\right)^2
+ g'\left(N\left(R\right)\right) g\left(N\left(R\right)\right)\right) \frac{dF(R)}{dR}
 - \frac{F(R)}{2}
+ \sum_i \rho_{i0} a_0^{-3(1+w_i)} \e^{-3(1+w_i)N(R)}\ ,
\eea
which constitutes a differential equation for $F(R)$, where the variable
is scalar curvature $R$.
Instead of $g$, if we use $G(N) \equiv g\left(N\right)^2 = H^2$,
the expression (\ref{RZ9}) could be a little bit simplified:
\bea
\label{RZ11}
0 &=& -9 G\left(N\left(R\right)\right)\left(4 G'\left(N\left(R\right)\right)
+ G''\left(N\left(R\right)\right)\right) \frac{d^2 F(R)}{dR^2}
+ \left( 3 G\left(N\left(R\right)\right)
+ \frac{3}{2} G'\left(N\left(R\right)\right) \right) \frac{dF(R)}{dR} \nn
&& - \frac{F(R)}{2}
+ \sum_i \rho_{i0} a_0^{-3(1+w_i)} \e^{-3(1+w_i)N(R)}\ .
\eea
Note that the scalar curvature is given by $R= 3 G'(N) + 12 G(N)$.
Hence, when we find $F(R)$ satisfying the differential equation
(\ref{RZ9}) or (\ref{RZ11}),
such $F(R)$ theory
admits the solution (\ref{RZ7}). Hence, such $F(R)$ gravity realizes
above cosmological solution.

As an example, we reconstruct the $F(R)$ gravity which reproduces the $\Lambda$CDM-era
but without real matter.
In the Einstein gravity, the FRW equation for the $\Lambda$CDM cosmology
is given by
\be
\label{RZ13}
\frac{3}{\kappa^2} H^2 = \frac{3}{\kappa^2} H_0^2 + \rho_0 a^{-3}
= \frac{3}{\kappa^2} H_0^2 + \rho_0 a_0^{-3} \e^{-3N} \ .
\ee
Here $H_0$ and $\rho_0$ are constants. The first term in the r.h.s. corresponds to the
cosmological constant and the second term to the cold dark matter (CDM).
The (effective) cosmological constant $\Lambda$
in the present universe is given by $\Lambda = 12 H_0^2$.
Then one gets
\be
\label{RZ14}
G(N) = H_0^2 + \frac{\kappa^2}{3} \rho_0 a_0^{-3} \e^{-3N} \ ,
\ee
and $R = 3 G'(N) + 12 G(N) = 12 H_0^2 + \kappa^2\rho_0 a_0^{-3} \e^{-3N}$,
which can be solved with respect to $N$ as follows,
\be
\label{RZ16}
N = - \frac{1}{3}\ln \left(\frac{ \left(R - 12 H_0^2\right)}{\kappa^2 \rho_0 a_0^{-3}}\right)\ .
\ee
Eq.(\ref{RZ11}) takes the following form:
\be
\label{RZ17}
0=3\left(R - 9H_0^2\right)\left(R - 12H_0^2\right) \frac{d^2 F(R)}{d^2 R}
 - \left( \frac{1}{2} R - 9 H_0^2 \right) \frac{d F(R)}{dR} - \frac{1}{2} F(R)\ .
\ee
By changing the variable from $R$ to $x$ by $x=\frac{R}{3H_0^2} - 3$,
Eq.(\ref{RZ17}) reduces to the hypergeometric differential equation:
\be
\label{RZ19}
0=x(1-x)\frac{d^2 F}{dx^2} + \left(\gamma - \left(\alpha + \beta + 1\right)x\right)\frac{dF}{dx}
 - \alpha \beta F\ .
\ee
Here
\be
\label{RZ20}
\gamma = - \frac{1}{2}\ ,\alpha + \beta = - \frac{1}{6}\ ,\quad \alpha\beta = - \frac{1}{6}\ ,
\ee
Solution of (\ref{RZ19}) is given by Gauss' hypergeometric function $F(\alpha,\beta,\gamma;x)$:
\be
\label{RZ22}
F(x) = A F(\alpha,\beta,\gamma;x) + B x^{1-\gamma} F(\alpha - \gamma + 1, \beta - \gamma + 1,
2-\gamma;x)\ .
\ee
Here $A$ and $B$ are constant.
Thus, we demonstrated that modified $F(R)$ gravity may describe the
$\Lambda$CDM epoch without the need to introduce the effective
cosmological constant.

As an another example, we reconstruct $F(R)$ gravity reproducing the
system with non-phantom matter
and phantom matter in the Einstein gravity, whose FRW equation is given by
\be
\label{RZ23}
\frac{3}{\kappa^2} H^2 = \rho_q a^{-c} + \rho_p a^c\ .
\ee
Here $\rho_q$, $\rho_p$, and $c$ are positive constants.
When $a$ is small as in the early universe, the first term in the r.h.s.
dominates and it behaves
as the universe described by the Einstein gravity with a matter whose EoS
parameter is $w=-1 + c/3>-1$,
that is, non-phantom like. On the other hand, when $a$ is large as in the
late universe, the second
term dominates and behaves as a phantom-like matter with $w= - 1 - c/3 < -1$.
Then since $G(N) \equiv g\left(N\right)^2 = H^2$, we find
\be
\label{RZ24}
G = G_q \e^{-cN} + G_p \e^{cN}\ ,\quad G_q \equiv \frac{\kappa^2}{3}\rho_q a_0^{-c}\ ,\quad
\quad G_p \equiv \frac{\kappa^2}{3}\rho_p a_0^c\ .
\ee
Then since $R = 3 G'(N) + 12 G(N)$,
\be
\label{RX25}
\e^{cN}=\frac{R\pm \sqrt{R^2 - 4\left(144 - 9c^2\right)}}{2\left(12 + 3 c\right)}\ ,
\ee
when $c\neq 4$ and
\be
\label{RX26}
\e^{cN}=\frac{R}{24G_p}\ ,
\ee
when $c=4$.
In the following, just for simplicity, we consider $c=4$ case. In the case, the
non-phantom matter corresponding to the first term in the r.h.s. of
(\ref{RZ23}) could be radiation with $w=1/3$. Then Eq.(\ref{RZ11}) in this case
is given by
\be
\label{RX27}
0 = -6 \left(\frac{24 G_p G_q}{R} + \frac{R}{24}\right) R \frac{d^2 F(R)}{d R^2}
+ \frac{9}{2}\left( - \frac{24 G_p G_q}{R} + \frac{R}{24}\right) \frac{d F(R)}{d R}
 - \frac{F(R)}{2}\ .
\ee
By changing variable $R$ to $x$ by $R^2 = - 576 G_p G_q x$,
we can rewrite Eq.(\ref{RX27}) as
\be
\label{RX29}
0 = (1-x)x\frac{d^2 F}{dx^2} + \left( \frac{3}{4} + \frac{x}{4}\right) \frac{dF}{dx}
 - \frac{F}{2}\ ,
\ee
whose solutions are again given by Gauss' hypergeometric function (\ref{RZ22}) with
\be
\label{RX30}
\gamma = \frac{3}{4}\ ,\quad \alpha + \beta + 1 = - \frac{1}{4}\ ,\quad
\alpha \beta = \frac{1}{2}\ .
\ee
Let us now study a model where the dominant component is phantom-like one.
Such kind of system can
be easily expressed in the standard General Relativity when a phantom
fluid is considered, where
the FRW equation reads $H^2(t)=\frac{\kappa^2}{3}\rho_\mathrm{ph}$,
Here the subscript $ph$ denotes the phantom nature of the fluid.
As the EoS for the fluid is given by $p_\mathrm{ph}=w_\mathrm{ph}\rho_\mathrm{ph}$
with $w_\mathrm{ph}<-1$,
by using the conservation equation $\dot{\rho}_\mathrm{ph}+3H(1+w_\mathrm{ph})\rho_\mathrm{ph}=0$,
the solution
for the FRW equation $H^2(t)=\frac{\kappa^2}{3}\rho_\mathrm{ph}$ is well known,
and it yields $a(t)=a_0(t_s-t)^{-H_0}$,
where $a_0$ is a constant, $H_0=-\frac{1}{3(1+w_\mathrm{ph})}$ and $t_s$ is the so-called Rip time.
Then, the solution describes the Universe that ends at the
Big Rip singularity in
the time $t_s$. The same behavior can be achieved in $F(R)$ theory with no need to introduce
a phantom fluid. The equation (\ref{RZ11}) can be solved and the expression for the $F(R)$ that
reproduces the solution is reconstructed. The expression for the Hubble parameter
as a function of the number of e-folds is given by
$H^2(N)=H^2_0 \e^{2N/H_0}$.
Then, the equation (\ref{RZ11}), with no matter contribution, takes the form:
\be
R^2\frac{d^2F(R)}{dR^2}+AR\frac{dF(R)}{dR}+BF(R)=0\ ,
\label{RX34}
\ee
where $A=-H_0(1+H_0)$ and $B=\frac{(1+2H_0)}{2}$. This equation is the well known Euler
equation whose solution yields
\be
F(R)=C_1R^{m_+}+C_2R^{m_-}\ , \quad \text{where} \quad m_{\pm}=\frac{1-A\pm\sqrt{(A-1)^2-4B}}{2}\ .
\label{RX35}
\ee
Thus, the phantom dark energy cosmology $a(t)=a_0(t_s-t)^{-H_0}$ can be
also obtained in the frame of $F(R)$ theory and no phantom fluid is needed.

We can consider now the model where the transition to the phantom epoch
occurs. It has been
pointed out that $F(R)$ could behave as an effective cosmological
constant, such that its current
observed value is well reproduced. One can reconstruct the model where
late-time acceleration is reproduced
by an effective cosmological constant and then the phantom barrier is
crossed (see ref.\cite{bamba} for such reconstruction in the presence of
auxiliary
scalar). Such transition,
which may take place at current time, could be achieved in $F(R)$
gravity. The solution considered can be expressed as:
\be
H^2=H_1\left( \frac{a}{a_0}\right)^{m}+H_0=H_1 \e^{m N}+H_0\ ,
\label{RX36}
\ee
where $H_1$,$H_0$ and $\alpha$ are positive constants. This solution can be constructed
in GR when a cosmological constant and a phantom fluid are included. In
the present case,
the solution (\ref{RX36}) can be achieved just by an $F(R)$ function, such that the transition
from non-phantom to phantom epoch is reproduced.
Scalar curvature can be written in terms of the number of e-folds again.
Then, the equation (\ref{RZ11}) takes the form:
\be
x(1-x)F''(x)+\left[x\left( -\frac{6+m}{6m}\right) -\frac{1}{3m} \right] F'(x)-\frac{m +4}{m}F(x)=0\ ,
\label{RX38}
\ee
where $x=\frac{1}{3H_0(m+4)}(12H_0-R)$.
The equation (\ref{RX38}) reduces to the
hypergeometric differential equation (\ref{RZ22}), so the solution is given, as in some
of the examples studied above, by the Gauss' hypergeometric function (\ref{RZ23}), whose parameters
for this case are given by
\be
\gamma= -\frac{1}{3m}\ , \quad \alpha+\beta=-\frac{3m+2}{2m}\ , \quad \alpha\beta=\frac{m+4}{2m}\ ,
\label{RX39}
\ee
and the obtained $F(R)$ gravity produces the FRW cosmology with the
late-time crossing of the phantom
barrier in the universe evolution.

Another example with transient phantom behavior in $F(R)$ gravity can be
achieved by following the same
reconstruction described above. In this case, we consider the following Hubble parameter:
\be
H^2(N)=H_0\ln\left(\frac{a}{a_0} \right)+H_1=H_0N+H_1\ ,
\label{RX40}
\ee
where $H_0$ and $H_1$ are positive constants. For this model, we have a contribution of
an effective cosmological constant, and a term that will produce
a superaccelerating phase
although no future singularity will take place(compare with
earlier model \cite{abdalla} with transient phantom era). The solution for
the model (\ref{RX40})
can be expressed as a function of time
\bea
H(t)=\frac{a_0H_0}{2}(t-t_0) \
\label{RX41}
\eea
Then, the Universe is superaccelerating, but as it can be seen from
(\ref{RX41}), in spite of its phantom nature,
no future singularity occurs.
The differential reconstruction equation can be obtained as
\be
a_2x\frac{d^2F(x)}{dx}+(a_1x+b_1)\frac{dF(x)}{dx}+b_0F(x)=0\ ,
\label{RX44}
\ee
where we have performed a variable change $x=H_0N+H_1$, and the constant parameters are
$a_2=H^2_0$, $a_1=-H_0$ $b_1=-\frac{H^2_0}{2}$ and $b_0=2H_0$. The equation (\ref{RX44}) is
a kind of the degenerate hypergeometric equation, whose solutions are given by the Kummers
series $K(a,b;x)$:
\be
F(R)=K\left(-2, -\frac{1}{2H_0};\frac{R-3H_0}{12} \right)\ .
\label{RX45}
\ee
Hence, such $F(R)$ gravity has cosmological solution with the transient
phantom behavior which does not evolve to future singularity.

Let us now consider the case where a future contracting Universe is reconstructed in this kind
of models. We study a model where the universe is currently accelerating,
then the future contraction
of the Universe occurs. The following solution for the Hubble parameter is considered,
\be
H(t)=2H_1(t_0-t)\ ,
\label{RS1}
\ee
where $H_1$ and $t_0$ are positive constants. For this example, the Hubble parameter (\ref{RS1})
turns negative for $t>t_0$, when the Universe starts to contract itself, while for $t\ll t_0$,
the cosmology is typically $\Lambda$CDM one.
Using notations $\tilde{H_0}=4H_1 t_0^2 $ and $\tilde{H_1}=4H_1$ and
repeating the above calculation, one gets:
\be
F(R)=K\left(-8\tilde{H_1}, -\frac{\tilde{H_1}}{8}; \frac{12\tilde{H_0}-3\tilde{H_1}-R}{12\tilde{H_1}} \right)\ .
\label{RS7}
\ee
Hence, the oscillating cosmology (\ref{RS1}) that describes the
asymptotically contracting Universe with a current
accelerated epoch can be found in specific $F(R)$ gravity.

Thus, we explicitly demonstrated that $F(R)$ gravity reconstruction is
possible for any cosmology under consideration without the need to
introduce the auxiliary scalar. However, the obtained
modified gravity has typically polynomial structure with terms which
contain positive and
negative powers of curvature as in the first such model unifying
the early-time inflation and late-time acceleration \cite{noprd}.
As a rule such models do not pass all the local gravitational tests.
Some generalization of above cosmological reconstruction is necessary.

\section{Cosmological reconstruction of viable $F(R)$ gravity}

In this section, we show how the cosmological reconstruction may be applied
to viable modified gravity which passes the local gravitational tests.
In this way, the non-linear structure of modified $F(R)$ gravity may be
accounted for, unlike the previous section where only polynomial $F(R)$
structures may be reconstructed.
Let us write $F(R)$ (\ref{Hm0}) in the following form:
$F(R) = F_0(R) + F_1(R)$.
Here we choose $F_0(R)$ as a known function like that of GR or one of
viable $F(R)$ models introduced in \cite{Hu:2007nk},
or viable $F(R)$ theories unifying inflation with dark 
energy \cite{Nojiri:2007as,Cognola:2007zu}, for example
\be
\label{RR4}
F_0(R) = \frac{1}{2\kappa^2}\left( R - \frac{\left(R-R_0\right)^{2n+1} + R_0^{2n+1}}{f_0
+ f_1 \left\{\left(R-R_0\right)^{2n+1} + R_0^{2n+1}\right\}}\right)\ .
\ee

Using the procedure similar to the one of second section, one gets the
reconstruction equation corresponding to (\ref{RZ11})
\bea
\label{RR5}
0 &=& -9 G\left(N\left(R\right)\right)\left(4 G'\left(N\left(R\right)\right)
+ G''\left(N\left(R\right)\right)\right) \frac{d^2 F_0(R)}{dR^2}
+ \left( 3 G\left(N\left(R\right)\right)
+ \frac{3}{2} G'\left(N\left(R\right)\right) \right) \frac{dF_0(R)}{dR} \nn
&& - \frac{F_0(R)}{2} \nn
&& -9 G\left(N\left(R\right)\right)\left(4 G'\left(N\left(R\right)\right)
+ G''\left(N\left(R\right)\right)\right) \frac{d^2 F_1(R)}{dR^2}
+ \left( 3 G\left(N\left(R\right)\right)
+ \frac{3}{2} G'\left(N\left(R\right)\right) \right) \frac{dF_1(R)}{dR} \nn
&& - \frac{F_1(R)}{2}
+ \sum_i \rho_{i0} a_0^{-3(1+w_i)} \e^{-3(1+w_i)N(R)}\ .
\eea
The above equation can be regarded as a differential equation for $F_1(R)$.
For a given $G(N)$ or $g(N)$ (\ref{RZ7}), if one can solve
(\ref{RZ11})) as $F(R)=\hat F(R)$, we also find the solution of
(\ref{RR5}) as $F_1(R) = \hat F (R) - F_0(R)$.
For example, for $G(N)$ (\ref{RZ14}), by using (\ref{RZ22}), we find
\be
\label{RR7}
F_1(R) = A F(\alpha,\beta,\gamma;x) + B x^{1-\gamma} F(\alpha - \gamma + 1, \beta - \gamma + 1,
2-\gamma;x) - F_0(R)\ .
\ee
Here $\alpha$, $\beta$, $\gamma$, and $x$ are given by $x=\frac{R}{3H_0^2} - 3$ and (\ref{RZ20}).
Using $F_0(R)$ (\ref{RR4}) one has
\be
\label{RR8}
F_1(R) = A F(\alpha,\beta,\gamma;x) + B x^{1-\gamma} F(\alpha - \gamma + 1, \beta - \gamma + 1,
2-\gamma;x) - \frac{1}{2\kappa^2}\left( R - \frac{\left(R-R_0\right)^{2n+1} + R_0^{2n+1}}{f_0
+ f_1 \left\{\left(R-R_0\right)^{2n+1} + R_0^{2n+1}\right\}}\right)\ ,
\ee
which describes the asymptotically de Sitter universe.
Instead of $x=\frac{R}{3H_0^2} - 3$ and (\ref{RZ20}), if we choose
$\alpha$, $\beta$, $\gamma$, and $x$ as $R^2 = - 576 G_p G_q x$ and in
(\ref{RX30}), $F_1(R)$ (\ref{RR8}) shows the
asymptotically phantom universe behavior, where
$H$ diverges in future.

One may start from $F_0(R)$ given by hypergeometric function (\ref{RZ22})
with $x=\frac{1}{3H_0(m+4)}(12H_0-R)$ and (\ref{RX39}).
In such a model, there occurs Big Rip singularity.
Let $\tilde F(R)$ be $F(R)$ again given by hypergeometric function (\ref{RZ22}) with
$x=\frac{R}{3H_0^2} - 3$ and (\ref{RZ20}):
\bea
\label{RRR2}
&& \tilde F(R) = \tilde A F(\tilde\alpha, \tilde\beta, \tilde\gamma; \tilde x)
+ \tilde B \tilde x^{1 - \tilde\gamma} F(\tilde\alpha - \tilde\gamma + 1, \tilde\beta
 - \tilde\gamma + 1, 2 - \tilde\gamma ; \tilde x)\ , \nn
&& \tilde x=\frac{R}{3H_0^2} - 3\ ,\quad
\tilde \gamma = - \frac{1}{2}\ , \tilde\alpha + \tilde \beta = - \frac{1}{6}\ ,
\quad \tilde\alpha \tilde\beta = - \frac{1}{6}\ .
\eea
If we choose $F(R)=\tilde F(R)$, the $\Lambda$CDM model emerges.
Then choosing $F_1(R) = \tilde F(R) - F_0(R)$,
the Big Rip singularity, which occurs in $F_0(R)$ model, does not appear
and
the universe becomes asymptotically de Sitter space. Hence, the
reconstruction method suggests the way to create the non-singular modified
gravity models \cite{abdalla,bamba,nonsingular}. Of course, it should be
checked that reconstruction term is not large (or it affects only the
very early-time/late-time universe) so that the theory passes the local
tests as it was before the adding of correction term.

Gauss' hypergeometric function $F(\alpha,\beta,\gamma;x)$ is defined by
\be
\label{RZ25}
F(\alpha,\beta,\gamma;x) = \frac{\Gamma(\gamma)}{\Gamma(\alpha) \Gamma(\beta)}\sum_{n=0}^\infty
\frac{\Gamma(\alpha+n) \Gamma(\beta+n)}{\Gamma(\gamma+n) n!}z^n\ .
\ee
Since
\be
\label{RZ27}
\alpha_0,\beta_0 = \frac{- 3m - 2 \pm \sqrt{m^2 - 20m +4}}{4m}<0\ ,\quad
\tilde\alpha, \tilde\beta = \frac{ -1 \pm 5 }{12}\ ,
\ee
when $R$ is large, $F_1(R)$ behaves as
$F_1(R) \sim R^{\left(3m+2 + \sqrt{m^2 - 20m +4}\right)/4m}$.
In spite of the above expression, since the total $F(R)=F_0(R) + F_1(R)$ is
given by $\tilde F(R)$ (\ref{RRR2}), the Big Rip type singularity does
not occur.
The asymptotic behavior of $F_1(R)$ cancels the large $R$ behavior in
$F_0(R)$ suggesting the way to present the non-singular cosmological
evolution.

We now consider the case that $H$ and therefore $G$ oscillate as
\be
\label{RR9b}
G(N) = G_0 + G_1 \sin \left( \frac{N}{N_0} \right)\ ,
\ee
with positive constants $G_0$, $G_1$, and $N_0$.
Let the amplitude of the oscillation is small but the frequency is large:
\be
\label{RR10}
G_0 \gg \frac{G_1}{N_0}\ ,\quad N_0 \gg 1\ .
\ee
When $G_1=0$, we obtain de Sitter space, where the scalar curvature is a constant $R=12 G_0$.
Writing $G(N)$ as
\be
\label{RR11}
G = \frac{R}{6} - G_0\ ,
\ee
by using (\ref{RZ11}), one arrives at general relativity:
\be
\label{RR12}
F(R) = c_0 \left( R - 6G_0 \right)\ .
\ee
Instead of (\ref{RR11}), using an arbitrary function $\tilde F$, if we write
\be
\label{RR12B}
G = G_0 + \tilde F (R) - \tilde F(12 G_0)\ ,
\ee
we obtain a general $F(R)$ gravity, which admits de Sitter space solution.
When $G_1 \neq 0$, under the assumption (\ref{RR10}), one may identify
$F(R)$ in (\ref{RR12}) with $F_0(R)$.
We now write $G(N)$ and the scalar curvature $R$ as
\be
\label{RR13}
G(N) = \frac{R}{6} - G_0 + \frac{G_1}{N_0}g(N)\ ,\quad R= 12 G_0 + \frac{3G_1}{N_0} r(N) \ ,
\ee
with adequate functions $g(N)$ and $r(N)$.
Then since $R = 6 g'(N) g(N) + 12 g(N)^2$ and from (\ref{RR10}), we find
\be
\label{RR14}
g(N) = - \left(N_0 \sin\frac{N}{N_0}+\frac{1}{2}\cos\frac{N}{N_0} \right)\ ,
\quad r(N)= 4N_0\sin\frac{N}{N_0}+\cos\frac{N}{N_0}
\ee
By assuming
\be
\label{RR15}
F(R) = c_0 \left( R - 6G_0 + \frac{G_1^2}{N_0^3} f(R) \right)\ ,
\ee
and identifying
\be
\label{RR16}
F_1(R) = \frac{c_0 G_1^2}{N_0^3} f(R)\ ,
\ee
from (\ref{RR5}), one obtains
\be
\label{RR17}
0 = G_0 \frac{df}{dr} - \sin \left(\frac{N}{N_0}\right)
+ o\left(\frac{G_1}{N_0},N_0\right)\ ,
\ee
which can be solved as
\be
\label{RR18}
f(R) = - \frac{1}{2G_0} \left(\cos^{-1} r \mp r \sqrt{1 - r^2 } \right)\ .
\ee
Then at least perturbatively, one can construct a model which exhibits the
oscillation of $H$.

Before going further, let us find $F(R)$ equivalent to the Einstein
gravity with a perfect fluid with
a constant EoS parameter $w$, where $H$ behaves as
\be
\label{RR19}
\frac{3}{\kappa^2} H^2 = \rho_0 \e^{-3(w+1)}\ .
\ee
Then
\be
\label{RR20}
G(N) = \frac{\kappa^2 \rho_0}{3}\e^{-3(w+1)}\ ,\quad
R(N) = \left(1 - 3w\right) \kappa^2 \rho_0 \e^{-3(w+1)}\ ,
\ee
which could be solved as
\be
\label{RR21}
N = - \frac{1}{3(w+1)} \ln \frac{R}{\left(1 - 3w\right) \kappa^2 \rho_0 }\ .
\ee
Therefore Eq.(\ref{RZ11}) has the following form:
\be
\label{RR22}
0= \frac{3(1+w)}{1-3w}R^2 \frac{d^2 F(R) }{dR^2} - \frac{1+3w}{2(1-3w)} R \frac{dF(R)}{dR}
 - \frac{F(R)}{2}\ ,
\ee
whose solutions are given by a sum of powers of $R$
\be
\label{RR23}
F(R)=F_+ R^{n_+} + F_- R^{n_-} \ .
\ee
Here $F_\pm$ are constants of integration and $n_\pm$ are given by
\be
\label{RR24}
n_\pm = \frac{1}{2} \left\{
\frac{7 + 9w}{6(1+w)} \pm \sqrt{
\left( \frac{7 + 9w}{6(1+w)} \right)^2 + \frac{2(1-3w)}{3(1+w)}
}\right\}\ .
\ee
If $w>-1/3$, the universe is decelerating but if $-1<w<-1/3$, the universe is accelerating
as in the quintessence scenario.

By using the solution (\ref{RZ22}), which mimics $\Lambda$CDM model,
 and the solution
(\ref{RR23}), one may consider the following model:
\be
\label{RR25}
F(x) = \left\{ A F(\alpha,\beta,\gamma;x) + B x^{1-\gamma} F(\alpha - \gamma + 1, \beta - \gamma + 1,
2-\gamma;x)\right\} \frac{\e^{\lambda\left(\frac{R}{R_1} - \frac{R_1}{R}\right)}}
{\e^{\lambda\left(\frac{R}{R_1} - \frac{R_1}{R}\right)}
+ \e^{- \lambda\left(\frac{R}{R_1} - \frac{R_1}{R}\right)}}
+ F_+ R^{n_+} + F_- R^{n_-}\ .
\ee
Here $R_1$ is a constant which is sufficiently small compared with the curvature $R_0$ in the
present universe. On the other hand, we choose a positive constant $\lambda$ to be large enough.
We also choose $F_\pm$ to be small enough so that only the first term dominates when $R\gg R_1$.
Note that the factor $\frac{\e^{\lambda\left(\frac{R}{R_1} -
\frac{R_1}{R}\right)}}
{\e^{\lambda\left(\frac{R}{R_1} - \frac{R_1}{R}\right)}
+ \e^{- \lambda\left(\frac{R}{R_1} - \frac{R_1}{R}\right)}}$ behaves as step function when
$\lambda$ is large:
\be
\label{RR26}
\lim_{\lambda\to + \infty}
\frac{\e^{\lambda\left(\frac{R}{R_1} - \frac{R_1}{R}\right)}}
{\e^{\lambda\left(\frac{R}{R_1} - \frac{R_1}{R}\right)}
+ \e^{- \lambda\left(\frac{R}{R_1} - \frac{R_1}{R}\right)}}
= \theta(R - R_1) \equiv \left\{
\begin{array}{ll}
1 & \mbox{when}\ R>R_1 \\
0 & \mbox{when}\ R<R_1 \\
\end{array}\right. \ .
\ee
Then in the early universe and in the present universe, only the first
term dominates and
the $\Lambda$CDM universe could be reproduced. In the future universe where $R\ll R_1$, the factor
$\frac{\e^{\lambda\left(\frac{R}{R_1} - 1\right)}}
{\e^{\lambda\left(\frac{R}{R_1} - 1\right)} + \e^{- \lambda\left(\frac{R}{R_1} - 1\right)}}$
decreases very rapidly and the second terms in
(\ref{RR25}) dominate. Then if $w>-1/3$, the universe decelerates again
but if $-1<w<-1/3$, the universe will be accelerating as
 in the quintessence scenario.

Thus, we explicitly demonstrated that the viable $F(R)$ gravity may be
reconstructed so that any requested cosmology may be realized after the
reconstruction. Moreover, one can use the viable $F(R)$ gravity unifying
the early-time inflation with late-time acceleration (and manifesting the
radiation/matter dominance era between accelerations) and passing local
tests in such a scheme. The (small) correction term $F_1(R)$ can be always
constructed so that it slightly corrects (if necessary) the cosmological bounds being
relevant only at the very early/late universe. This scenario opens the way
to extremely realistic description of the universe evolution in $F(R)$
gravity consistent with local tests and cosmological bounds.

\section{Reconstruction of modified gravity with extra scalar}

We now consider the reconstruction of $F(R)$-gravity coupled with a scalar field,
whose action is given by
\be
\label{FS1}
S= \int d^4x \sqrt{-g} \left(\frac{F(R)}{2\kappa^2}
 - \frac{1}{2}\partial_\mu \phi \partial^\mu \phi - V(\phi)
+ \mathcal{L}_\mathrm{matter} \right)\ .
\ee
Let us redefine the scalar field as $\phi=\phi(\varphi)$,
\be
\label{FS2}
S= \int d^4x \sqrt{-g} \left(\frac{F(R)}{2\kappa^2}
 - \frac{\omega(\varphi) }{2}\partial_\mu \varphi \partial^\mu \varphi - \tilde V(\varphi)
+ \mathcal{L}_\mathrm{matter} \right)\ .
\ee
Here
\be
\label{FS3}
\omega(\varphi) \equiv \left(\frac{d\phi(\varphi)}{d\varphi}\right)^2\ ,\quad
\tilde V(\varphi) \equiv V\left(\phi\left(\varphi\right)\right)\ .
\ee
If $\phi$ only depends on the time-coordinate $t$ or e-folding $N$, we may choose
$\varphi = t$ or $\varphi = N$.

Then the equations corresponding to the first and second FRW equations have the following form
\bea
\label{FS4}
0 &=& - \frac{F(R)}{2} + 3\left( H^2 + H H' \right) F'(R)
 - 18 \left( 4H^3 H' + H^2 \left(H'\right)^2 + H^3 H'' \right) F''(R) \nn
&& + \kappa^2 \left( \frac{H^2 \omega(\varphi) \left( \varphi' \right)^2}{2} + \tilde V(\varphi) \right)
+ \sum_i \kappa^2 \rho_{i0} a_0^{-3(1+w_i)} \e^{-3(1+w_i)N}\ , \\
\label{FS5}
0 &=& \frac{F(R)}{2} - \left( 3H^2 + HH' \right) F'(R)
+ 6 \left( 16 H^3 H' - 4H^2 \left(H'\right)^2 - H \left( H' \right)^3
 - 4 H^2 H' H'' - H^3 H''' \right) F''(R) \nn
&& - 36 \left( 4 H^3 H' + H^2 \left(H'\right)^2 + H^3 H'' \right)
\left( \left(H'\right)^2 + H H'' + 4 H H' \right) F'''(R) \nn
&& + \kappa^2 \left( \frac{H^2 \omega(\varphi) \left( \varphi' \right)^2}{2} - \tilde V(\varphi) \right)
+ \sum_i \kappa^2 w_i \rho_{i0} a_0^{-3(1+w_i)} \e^{-3(1+w_i)N}\ ,
\eea
which can be rewritten as
\bea
\label{FS6}
\kappa^2 \omega(\varphi)\left( \varphi' \right)^2
&=& \left\{ - 2HH' F'(R) + 6 \left( - 4 H^3 H' + 7 H^2 \left(H'\right)^2
+ H \left(H'\right)^3 + 4 H^2 H' H'' + H^3 H''' + 3 H^3 H'' \right) F''(R) \right. \nn
&& - 36 \left( 4 H^3 H' + H^2 \left(H'\right)^2 + H^3 H'' \right)
\left(\left(H'\right)^2 + H H'' + 4 H H' \right) F'''(R) \nn
&& \left. + \sum_i \kappa^2 \left(1 + w_i\right) \rho_{i0} a_0^{-3(1+w_i)} \e^{-3(1+w_i)N}
\right\}\frac{1}{H^2} \ ,\\
\label{FS7}
2\kappa^2 \tilde V(\varphi) &=& F(R) + \left( - 6 H^2 - 4H H' \right) F'(R) \nn
&& + 6 \left( 28 H^3 H' - H^2 \left(H'\right)^2 - H \left(H'\right)^3
 - 4 H^2 H' H'' + H^3 H''' - 3 H^3 H''\right) F''(R) \nn
&& + 36 \left( 4 H^3 H' + H^2 \left(H'\right)^2 + H^3 H'' \right)
\left( \left(H'\right)^2 + H H'' + 4 H H' \right) F'''(R) \nn
&& + \sum_i \kappa^2 \left(1 - w_i\right) \rho_{i0} a_0^{-3(1+w_i)} \e^{-3(1+w_i)N}\ .
\eea
Then if we consider the model given by an adequate function $K=K(\varphi)$,
\bea
\label{FS8}
\kappa^2 \omega(\varphi) &=& \left\{ - 2K(\varphi)K'(\varphi)
F'\left( 6K(\varphi)K'(\varphi) + 12 K(\varphi)^2\right) \right. \nn
&& + 6 \left( - 4 K(\varphi)^3 K'(\varphi) + 7 K(\varphi)^2 K'(\varphi)^2
+ K(\varphi) K'(\varphi)^3 + 4 K(\varphi)^2 K'(\varphi) K''(\varphi)
+ K(\varphi)^3 K'''(\varphi) \right. \nn
&& \left. + 3 K(\varphi)^3 K''(\varphi) \right)
F''\left(6K(\varphi)K'(\varphi) + 12 K(\varphi)^2\right) \nn
&& - 36 \left( 4 K(\varphi)^3 K'(\varphi)
+ K(\varphi)^2 K'(\varphi)^2 + K(\varphi)^3 K''(\varphi) \right)
\left( K'(\varphi)^2 + K(\varphi) K''(\varphi)
+ 4 K(\varphi) K'(\varphi) \right) \nn
&& \left. \times F'''\left(6K(\varphi)K'(\varphi) + 12 K(\varphi)^2\right)
+ \sum_i \kappa^2 \left(1 + w_i\right) \rho_{i0} a_0^{-3(1+w_i)} \e^{-3(1+w_i)\varphi}\right\}
\frac{1}{K(\varphi)^2}\ ,\\
\label{FS9}
2\kappa^2 \tilde V(\varphi) &=& F\left(6K(\varphi)K'(\varphi) + 12 K(\varphi)^2\right)
+ \left( - 6 K(\varphi)^2 - 4K(\varphi) K'(\varphi) \right)
F'\left( 6 K(\varphi)K'(\varphi) + 12 K(\varphi)^2\right) \nn
&& + 6 \left( 28 K(\varphi)^3 K'(\varphi) - K(\varphi)^2 K'(\varphi)^2
 - K(\varphi) K'(\varphi)^3 - 4 K(\varphi)^2 K'(\varphi) K''(\varphi)
+ K(\varphi)^3 K'''(\varphi) \right. \nn
&& \left. - 3 K(\varphi)^3 K''(\varphi)\right)
F''\left( 6 K(\varphi)K'(\varphi) + 12 K(\varphi)^2\right) \nn
&& + 36 \left( 4 K(\varphi)^3 K'(\varphi) + K(\varphi)^2 K'(\varphi)^2
+ K(\varphi)^3 K''(\varphi) \right)
 \left( K'(\varphi)^2 + K(\varphi) K''(\varphi) + 4 K(\varphi) K'(\varphi) \right) \nn
&& \times F'''\left( 6 K(\varphi)K'(\varphi) + 12 K(\varphi)^2\right)
+ \sum_i \kappa^2 \left(1 - w_i\right) \rho_{i0} a_0^{-3(1+w_i)} \e^{-3(1+w_i)\varphi}\ ,
\eea
we find a solution is given by
\be
\label{FS10}
H (N) = K (N)\ ,\quad \varphi = N\ .
\ee
Hence, it is demonstrated that reconstruction can be extended to the case
when modified gravity couples with some scalar field. Note that extra
scalar may be necessary in the situation when some of cosmological bounds
(for instance, cosmological perturbations theory which is extremely
complicated in $F(R)$ gravity, for a review, see \cite{sante}) cannot be
passed within only modified gravity. So far, this is not the case and
modified gravity which passes local tests and cosmological bounds is
available \cite{Hu:2007nk,Nojiri:2007as,Cognola:2007zu}.

\section{Discussion}

In summary, we developed general scheme for cosmological reconstruction of
modified $F(R)$ gravity in terms of e-folding (or redshift) without use of
auxiliary scalar in intermediate calculations. Using this method, it is
possible to construct the specific modified gravity which contains any
requested FRW cosmology. The number of $F(R)$ gravity examples is
found where the
following background evolution may be realized:$\Lambda$CDM epoch,
deceleration with subsequent transition to effective phantom
superacceleration leading to Big Rip singularity, deceleration with
transition to transient phantom phase without future singularity,
oscillating universe. It is important that all these cosmologies may be
realized only by modified gravity without use of any dark components
(cosmological constant, phantom, quintessence, etc).

It is shown that our method may be applied to viable $F(R)$ gravities
which
pass local tests and unify the early-time inflation with late-time
acceleration. In this case, the additional reconstruction may be made so
that correction term is not large and it is relevant only in the very
early/very late universe. Hence, the purpose of such additional
reconstruction is only to improve the cosmological predictions if the
original theory does not pass correctly the precise observational
cosmological
bounds. For instance, in this way it is possible to formulate the modified
gravity without finite-time future singularity. It is also demonstrated
that the reconstruction scheme may be generalized for the case of modified
gravity with scalar field.

The present reconstruction formulation shows that even if specific
realistic modified gravity
does not pass correctly some cosmological bounds (for instance, does not
lead to correct cosmological perturbations structure) it may be improved
with eventually desirable result. Hence, the successful development of
such method adds very strong argument in
favour of unified gravitational alternative for inflation, dark energy and
dark matter.

\section*{Acknowledgments}

The work by S.N. is supported in part by Global
COE Program of Nagoya University provided by the Japan Society
for the Promotion of Science (G07).
The work by S.D.O. is supported in part by MICINN (Spain) project
FIS2006-02842, by AGAUR (Generalitat de Catalunya), project 2009 SGR994
and by LRSS project N.2553.2008.2. DSG acknowledges a grant from MICINN.


\begin{thebibliography}{99}

\bibitem{review}
S.~Nojiri and S.~D.~Odintsov,
  eConf {\bf C0602061}, 06 (2006)
  [Int.\ J.\ Geom.\ Meth.\ Mod.\ Phys.\  {\bf 4}, 115 (2007)]
  [arXiv:hep-th/0601213]; [arXiv:0807.0685];
S.~Capozziello and M.~Francaviglia,
  Gen.\ Rel.\ Grav.\  {\bf 40}, 357 (2008)
  [arXiv:0706.1146 [astro-ph]];
T.~P.~Sotiriou and V.~Faraoni,
  arXiv:0805.1726 [gr-qc];
F.~S.~N.~Lobo,
  arXiv:0807.1640 [gr-qc].

\bibitem{reconstruction1}
S.~Nojiri and S.~D.~Odintsov,
  J.\ Phys.\ Conf.\ Ser.\  {\bf 66}, 012005 (2007)
  [arXiv:hep-th/0611071].

\bibitem{reconstruction2}
S.~Nojiri and S.~D.~Odintsov,
  Phys.\ Rev.\  D {\bf 74}, 086005 (2006)
  [arXiv:hep-th/0608008];
  J.\ Phys.\ A  {\bf 40}, 6725 (2007)
  [arXiv:hep-th/0610164];
S.~Capozziello, S.~Nojiri, S.~D.~Odintsov and A.~Troisi,
  Phys.\ Lett.\  B {\bf 639}, 135 (2006)
  [arXiv:astro-ph/0604431];
E.~Elizalde and D.~Saez-Gomez,
  arXiv:0903.2732 [hep-th].

\bibitem{reconstruction3}
A.~de la Cruz-Dombriz and A.~Dobado,
  Phys.\ Rev.\  D {\bf 74}, 087501 (2006)
  [arXiv:gr-qc/0607118];
J.~L.~Cortes and J.~Indurain,
  Astropart.\ Phys.\  {\bf 31}, 177 (2009)
  [arXiv:0805.3481 [astro-ph]];
I.~H.~Brevik,
  Gen.\ Rel.\ Grav.\  {\bf 38}, 1317 (2006)
  [arXiv:gr-qc/0603025];
L.~N.~Granda,
  arXiv:0812.1596 [hep-th];
M.~R.~Setare,
  Int.\ J.\ Mod.\ Phys.\  D {\bf 17}, 2219 (2008)
  [arXiv:0901.3252 [hep-th]];
X.~Wu and Z.~H.~Zhu,
  Phys.\ Lett.\  B {\bf 660}, 293 (2008)
  [arXiv:0712.3603 [astro-ph]].

\bibitem{bamba}
K.~Bamba, C.~Q.~Geng, S.~Nojiri and S.~D.~Odintsov,
  Phys.\ Rev.\  D {\bf 79}, 083014 (2009)
  [arXiv:0810.4296 [hep-th]];
K.~Bamba, S.~Nojiri and S.~D.~Odintsov,
  JCAP {\bf 0810}, 045 (2008)
  [arXiv:0807.2575 [hep-th]];
K.~Bamba and C.~Q.~Geng,
  arXiv:0901.1509 [hep-th].

\bibitem{abdalla}
M.~C.~B.~Abdalla, S.~Nojiri and S.~D.~Odintsov,
  Class.\ Quant.\ Grav.\  {\bf 22}, L35 (2005)
  [arXiv:hep-th/0409177].

\bibitem{noprd}
S.~Nojiri and S.~D.~Odintsov,
  Phys.\ Rev.\  D {\bf 68}, 123512 (2003)
  [arXiv:hep-th/0307288];
  Gen.\ Rel.\ Grav.\  {\bf 36}, 1765 (2004)
  [arXiv:hep-th/0308176].

\bibitem{Hu:2007nk}
  W.~Hu and I.~Sawicki,
  Phys.\ Rev.\  D {\bf 76}, 064004 (2007)
  [arXiv:0705.1158 [astro-ph]];
S.~A.~Appleby and R.~A.~Battye,
  Phys.\ Lett.\  B {\bf 654}, 7 (2007)
  [arXiv:0705.3199 [astro-ph]];
S.~Tsujikawa,
  Phys.\ Rev.\  D {\bf 77}, 023507 (2008)
  [arXiv:0709.1391 [astro-ph]];
S.~Capozziello and S.~Tsujikawa,
  Phys.\ Rev.\  D {\bf 77}, 107501 (2008)
  [arXiv:0712.2268 [gr-qc]];
B.~Li and J.~D.~Barrow,
  Phys.\ Rev.\  D {\bf 75}, 084010 (2007)
  [arXiv:gr-qc/0701111].

\bibitem{Nojiri:2007as}
  S.~Nojiri and S.~D.~Odintsov,
  Phys.\ Lett.\  B {\bf 657}, 238 (2007)
  [arXiv:0707.1941 [hep-th]];
  Phys.\ Rev.\  D {\bf 77}, 026007 (2008)
  [arXiv:0710.1738 [hep-th]].

\bibitem{Cognola:2007zu}
  G.~Cognola, E.~Elizalde, S.~Nojiri, S.~D.~Odintsov, L.~Sebastiani and S.~Zerbini,
  Phys.\ Rev.\  D {\bf 77}, 046009 (2008)
  [arXiv:0712.4017 [hep-th]].

\bibitem{nonsingular}
S.~Nojiri and S.~D.~Odintsov,
  Phys.\ Rev.\  D {\bf 78}, 046006 (2008)
  [arXiv:0804.3519 [hep-th]];
T.~Kobayashi and K.~i.~Maeda,
  Phys.\ Rev.\  D {\bf 79}, 024009 (2009)
  [arXiv:0810.5664 [astro-ph]];
A.~Dev, D.~Jain, S.~Jhingan, S.~Nojiri, M.~Sami and I.~Thongkool,
  Phys.\ Rev.\  D {\bf 78}, 083515 (2008)
  [arXiv:0807.3445 [hep-th]];
M.~Sami,
  arXiv:0904.3445 [hep-th];
S.~Capozziello, M.~De Laurentis, S.~Nojiri and S.~D.~Odintsov,
  Phys.\ Rev.\  D {\bf 79}, 124007 (2009)
  [arXiv:0903.2753 [hep-th]].

\bibitem{sante}
S.~Carloni, P.~K.~S.~Dunsby and A.~Troisi,
  arXiv:0906.1998 [gr-qc].



\end{thebibliography}
\end{document}